\documentstyle[a4]{article}

\title{Charge conservation and Equivalence principle}

\author{S. Landau
\thanks{Fellow of CONICET} 
\emph{F. C. Astron{\'o}micas y Geof{\'\i}sicas, U. N. de La 
Plata} \\
P. D. Sisterna \emph{Facultad de C. Exactas, U. N. de Mar 
del Plata} \\
H. Vucetich
\thanks{Member of CONICET} 
\emph{F. C. Astron{\'o}micas y Geof{\'\i}sicas, U. N. de La 
Plata}}
\begin{document}
\maketitle

\begin{abstract}
The $TH\epsilon\mu$ formalism was developed to study nonmetric
theories of gravitation. In this letter we show that theories that
violate Local Lorentz Invariance (LLI) or Local Position Invariance
(LPI) also violate charge conservation. Using upper bounds on this
violation we can put very stringent limits to violations of Einstein
Equivalence Principle (EEP). These limits, in turn, severely restrict
string-based models of low energy physics.
\end{abstract}

\begin{quote}
PACS: 4.80Cc, 11.30-J, 4.50+h
\end{quote}

There are two theoretical frameworks which stand as the milestones of
modern physics: the standard model of special relativistic particle
physics, and general relativity as the gravitational theory. The
former rests on gauge invariance symmetry, while the latter is built
geometrically from EEP. Two observational testable laws follow from
them: charge conservation (global gauge symmetry) and the weak
equivalence principle (invariance of non gravitational laws in locally
inertial frames).  The local aspects of both schemes can change
dramatically if there are long range interactions whose dynamics
cannot be influenced in local experiments. One can even expect that
local-frame global gauge invariance (not the same as local gauge
invariance) as well as EEP may be violated, even if the complete
(global) theory satisfy the invariances mentioned above.

Regarding possible violations of EEP, a scheme was developed at the
beginning of the seventies (\cite{Ni}) in order to analyze non metric
gravitational theories in spherically symmetric static
situations. This theoretical scheme, called the $TH\epsilon\mu$
formalism, has also been used to ``prove'' Schiff´s conjecture
\cite{Will93}.  By non metric we mean theories that present long range
fields (gravitation like fields) that couple with matter directly,
besides the metric (which may still account for part of the
gravitational sector of the theory). For instance, if there is a
scalar field with long range interactions that couple directly with
matter, then in a local falling frame, where the metric reduces to its
Minkowskian form, we may still have time- or space-dependent factors
in the local dynamics, which could render a non relativistic invariant
local lagrangian. Any ``external structure'' (e.g. Minkowski-metric
external structure replaced by the [dynamical] metric in general
relativity \cite{Norton93}) such as the fundamental constants can be
suspected of hiding long range fields that have frozen at some value,
making the fundamental parameters effectively constant.  Unification
schemes such as superstring theories \cite{Schwartz} and Kaluza-Klein
theories \cite{KK} have cosmological solutions in which the
low energy fundamental constants are functions of time (including
possibly the speed of light \cite{Moffat,AlbMag99}), thus violating
LLI and Local Position Invariance (LPI).

Usually low-energy phenomena are used to constrain the variation rate
of fundamental constants \cite{Ellis,Mc.Elhinny,Damour88,Hellings,%
Wolfe,Shlyakhter,SV1,SV2,Prestage,Kim}.  It is well known that objects
with space- or time-dependent masses follow paths which do not
correspond to geodesics of the space-time metric
\cite{Bekenstein77}. Indeed this violations of LPI induce violations
of the university of free fall, thus being subject to very stringent
tests. This tests are E\"otvos-type experiments, in which the
accelerations of neutral masses with different composition in the same
gravitational field are compared (null gravitational acceleration
experiments). These are the most precise tests of the equivalence
principle, reaching upper bounds of order $10^{-12}$
\cite{Su94,Will98} for the free fall parameter
$\eta(A,B)=(a(A)-a(B))/g$ , where $a(A)$ and $a(B)$ are the
accelerations of bodies $A$ and $B$ respectively and $g$ a local
reference gravitational acceleration.  In this letter we analyze the
local electromagnetic equations in the $TH\epsilon\mu$ formalism, and
show that there is an adiabatic non conservation of charge as measured
in local experiments.  We then analyze both a superstring
based and a Bekenstein-like model on which we put stringent upper
bounds on any violations of EEP several orders of magnitude tighter
than any previous one. In this way, we obtain an efective test for
string based and similar theories.

1: {\it Charge conservation in the $TH\epsilon\mu$ 
formalism:} \label{Q-THem}

Let $S_{NG}$ be the action defining the $TH\epsilon\mu$ formalism
\cite{Will93}:
\begin{eqnarray}
S_{NG} &=& - \sum_a m_{0a} \int(T - H v_a^2)^{1/2} dt \nonumber \\
& & + \sum_a e_a \int A_\mu v^\mu dt \nonumber\\
& & + \frac{1}{8\pi} \int(\epsilon E^2 - \mu^{-1} B^2) d^4 x
\label{S_NG}
\end{eqnarray}
where $T, H, \epsilon, \mu$ are functions of the spherically
symmetrical gravitational potential $\Phi(\mathbf{x})$.  It is assumed
that these functions are slowly varying in the neighborhood of a given
event $\mathcal{P}$, of the system $NG$.  We shall
choose a coordinate system with origin at $\mathcal{P}$ and approximate
these functions by linear functions of the local coordinates
within the volume $V$ of the system. So, 
in the neighborhood of $\mathcal{P}$ we can expand 
the gravitational potential in the form:
\begin{equation}
\Phi(\mathbf{r}) = \Phi_0 + \mathbf{f}_0 \cdot \mathbf{x} + \cdots
	\label{Phi-in-V}
\end{equation}
where $\mathbf{f}_0$ is proportional to the local acceleration of
gravity $\mathbf{g}_0$. In the same way we find
 $T_0 + {T_0}' \mathbf{f_0} \cdot
\mathbf{{x}}$, and similar expressions for $H$, $\epsilon$ and
$\mu$. Finally, we scale the coordinates in the form%
\footnote{This scaling is a particular case of a more general
transformation to a freely falling reference system, see references
\cite{Will93,GMSM}.}:
\begin{eqnarray} 
\hat{t} = T_0^{1/2} t & & \hat{\mathbf{x}} = H_0^{1/2}{\bf x}  
\label{scale0}
\end{eqnarray}

Then, in the neighborhood of event $\mathcal{P}$, 
the action takes the form:
\begin{eqnarray}
{S}_{NG} &=& -\sum_a m_{0a} \int (1 - \hat{v}^2)^{1/2} d \hat{t}
+ \sum_a e_a \int \hat{A}_{\hat{\mu}} \hat{v}^{\hat{\mu}} 
d\hat{t}
\nonumber \\
&+& \frac{1}{8\pi} \epsilon_0 {\left(\frac{T_0}{H_0}\right)}^{1/2} 
\int d^4 \hat{x}  \hat{E}^2 \epsilon^*(\hat{x})
 \nonumber \\
&-& \frac{1}{8\pi} {\left( \frac{H_0}{T_0}\right)}^{1/2}\mu_0^{-1} 
\int d^4 \hat{x}  \hat{B}^2 {\mu^*}^{-1}(\hat{x}) 
 \nonumber \\
&+& \frac{1}{8\pi} {\mathbf{f_0}} \cdot 
\int d^4\hat{x} \left( \hat{\mathbf{E}} \times \hat{\mathbf{B}}\right)
\sigma^*(\hat{x})
\label{S_NG-scaled}
\end{eqnarray}
where
\begin{eqnarray}
\epsilon^* &=& \left(1+\frac{1}{2}\frac{{T_0}'}{T_0}\frac{\Gamma_0}{H_0^
{1/2}}\mathbf{f_0} 
\cdot \mathbf{\hat{x}} \right)\\
{\mu^*}^{-1} &=& \left( 1- \frac{1}{2} \frac{{T_0}'}{T_0} 
\frac{\Lambda_0}{H_0^{1/2}} \mathbf{f_0} \cdot 
\mathbf{\hat{x}}\right)\\
\sigma^* &=& \epsilon_0 {T_0}' H_0^{1/2} \hat{t}\left(
1-\frac{H_0}{T_0}\epsilon_0^{-1} \mu_0^{-1}\right)
\end{eqnarray}
and
\begin{eqnarray}
\Gamma_0 = 2 \frac{T_0}{T'_0} \left(\frac{{\epsilon_0}'}{\epsilon_0}  +
\frac{1}{2} \frac{{T_0}'}{T_0} -
\frac{1}{2} \frac{{H_0}'}{H_0}
\right) \label{Def-Gamma-0} \\
\Lambda_0 = 2 \frac{T_0}{T'_0} \left(\frac{{\mu_0}'}{\mu_0} 
 +
\frac{1}{2} \frac{{T_0}'}{T_0} -
\frac{1}{2} \frac{{H_0}'}{H_0}
\right) \label{Def-Lambda-0} 
\end{eqnarray}
As usual, electric and magnetic fields are related to the local scalar
and vector potentials in the form:
\begin{equation}
\begin{array}{cc}
\hat{\mathbf{E}} = \hat{\nabla} \hat{A}_{\hat{0}} -
\hat{\mathbf{A}}_{,\hat{0}} &
\hat{\mathbf{B}} = \hat{\nabla} \times \hat{\mathbf{A}} 
\end{array} \label{Local-Pots}
\end{equation}

In equation \ref{S_NG-scaled}, we shall make a 
final scaling:
\begin{equation}
\begin{array}{cc}
e^*_a = T_0^{-1/4} H_0^{1/4} \epsilon_0^{-1/2} &
A^*_{\hat{\mu}} = T_0^{1/4} H_0^{-1/4} \epsilon_0^{1/2}
\end{array} \label{Local-Charge}
\end{equation}
which introduces the local particle charge $e^*_a$.  Besides we
introduce the local limiting velocity $c_0$, the local light velocity
$c_l$ and the ratio of both quantities $c^* = c_l/c_0$:
\begin{equation}
\begin{array}{ccc}
c_0 &=& (T_0/H_0)^{1/2} \nonumber \\
c_l &=& (\epsilon_0 \mu_0)^{-1/2} \nonumber \\
c^* &=& (T_0^{-1} H_0 \epsilon_0^{-1} \mu_0^{-1})^{1/2} \nonumber
\end{array} \label{Local-cs}
\end{equation}

\begin{eqnarray}
S_{NG} &=&  -\sum_a m_{0a} \int (1 - \hat{v}^2)^{1/2} d \hat{t}
 + \sum_a e_a^* \int \hat{A^*}_{\hat{\mu}} v^{\hat{\mu}} 
d\hat{t} \label{S_NG-renor}\\
 &+& \frac{1}{8\pi} \int d^4 \hat{x} \left[ \epsilon^* \hat{E^*}^2 - 
\mu^{*-1} \hat{B^*}^2 + \sigma^* \mathbf{f_0} \cdot(\hat{\mathbf{E}^*
}\times\hat{\mathbf{B}^*})
\right] \nonumber
\end{eqnarray}

Let us now introduce the local (renormalized) charge and current
density:
\begin{eqnarray}
\hat{\rho}^* &=& \sum_a e^*_a \delta\left(\mathbf{r} -
\mathbf{r}_a\right) \label{Def-rho-ren}\\
\hat{\mathbf{j}}^* &=& \sum_a e^*_a
\hat{\mathbf{v}}_a\delta\left(\mathbf{r} - \mathbf{r}_a\right)
\label{Def-j-ren}
\end{eqnarray}

Variation of (\ref{S_NG-renor}) yields the inhomogeneous pair of
Maxwell equations:
\begin{eqnarray}
\hat{\nabla} \cdot (\epsilon^* \mathbf{E}^*+\sigma^* 
\mathbf{B}^*\times \mathbf{g_0}) &=& 4\pi\rho^* \nonumber 
\\
\hat{\nabla} \times (\mu^{*-1} \mathbf{B}^*)& =& 
\frac{\partial}{\partial \hat t} (\epsilon^* \mathbf{E}^* +
\sigma^*\mathbf{B}^*\times \mathbf{g_0}) \nonumber \\
 & &+ \hat{\nabla} \times (\sigma^{*}\mathbf{g_0}\times
\mathbf{E}^*) \nonumber \\
& & + 4\pi \mathbf{j}^*.
\end{eqnarray}
It is apparent that local conservation of charge still 
holds, as it is easy 
to derive the equation:
\begin{equation}
\hat{\nabla} \cdot \mathbf{j}^* + \dot{\rho}^* = 0
\label{ConsChargeEq} 
\end{equation}

Thus, the locally conserved quantity is, in the $TH\epsilon\mu$
formalism:
\begin{equation}
Q^* = \int_V \rho^* d^3 \hat{x} \label{RenCharge}
\end{equation}
where the volume $V$ is small in comparison with the scale of
variation of the gravitational field $V^{1/3} \ll L_g$.  For a system
of identical particles, this may be written in the form:
\begin{equation}
Q^* = e^* N \label{RenCharge-Id}
\end{equation}

	Consider now an adiabatic change of $e^*$. Then the condition
$\dot{Q}^*=0$ implies:
\begin{equation}
\frac{\dot{N}}{N} = - \frac{\dot{e}^*}{e^*} = 
\frac{1}{2} \left( \frac{\dot{\epsilon_0}}{\epsilon_0} +
\frac{1}{2} \frac{\dot{T_0}}{T_0} -
\frac{1}{2} \frac{\dot{H_0}}{H_0}
\right) \label{Var-N-1}
\end{equation}
and, using $\dot{\epsilon} = \epsilon\dot{\Phi}$ and similar
expressions, we find:
\begin{equation}
\frac{\dot{N}}{N} =
\frac{1}{2} \left( \frac{{\epsilon_0}'}{\epsilon_0} 
+\frac{1}{2} \frac{{T_0}'}{T_0} -
\frac{1}{2} \frac{{H_0}'}{H_0}
\right) \dot{\Phi} \label{Var-N-2}
\end{equation}

	Now, from eq.\ref{Phi-in-V}: 
\begin{equation}
\dot{\Phi} = \mathbf{f}_0 \cdot \dot{\mathbf{x}} + \cdots
\label{dot-Phi}
\end{equation}
and $\mathbf{f_0}$ must be related to the local gravitational
acceleration which is defined as the acceleration of a structureless
particle. To find it, let us expand the first term of equation
(\ref{S_NG}) in the neighborhood of $\mathcal{P}$. For a single
uncharged particle we find:
\begin{equation}
S_{P} \simeq m_0 T_0^{1/2} \int \left( \frac{1}{2} 
\frac{v^2}{c_0^2}
- \frac{1}{2} \frac{T'_0}{T_0} \mathbf{f}_0 \cdot
{\mathbf{x}} \right) dt
\end{equation}
The corresponding equation of motion is:
\begin{equation}
\frac{1}{c_0^2} \frac{d^2 \mathbf{x}}{dt^2} =
\frac{1}{2} \frac{T'_0}{T_0} \mathbf{f}_0
\end{equation}
so we obtain, correct to the newtonian order:
\begin{equation}
{\mathbf{f}}_0 = \frac{T_0}{H_0^{1/2}} \frac{T_0}{T'_0}
\frac{\hat{\mathbf{g}}_0}{c_0^2} 
\label{f-as-g}
\end{equation}

Finally, substitution in (\ref{dot-Phi}) and in
(\ref{Var-N-2}) yields:
\begin{equation}
\frac{\dot{N}}{N} =\frac{1}{2} \frac{T_0}{T'_0} \left( 
\frac{{\epsilon_0}'}{\epsilon_0} +
\frac{1}{2} \frac{{T_0}'}{T_0} -
\frac{1}{2} \frac{{H_0}'}{H_0}
\right) \hat{\mathbf{g}}_0 \cdot \hat{\mathbf{v}}
\label{Var-N-3} 
\end{equation}
and introducing the local time (in seconds) through $c_0 
t^* = \hat{t}$ we obtain:
\begin{equation}
\frac{\dot{N}}{N} = \frac{\Gamma_0}{4} \frac{{\mathbf{g}}_0 \cdot
{\mathbf{v}}}{c_0^2} \label{Var-N-Def}
\end{equation}
where the parameter $\Gamma_0$, defined in eq.\ref{Def-Gamma-0},
characterizes anomalous accelerations and anomalous mass tensors
\cite{Will93}: 
\begin{eqnarray}
\delta m_P &=& 2 \Gamma_0 \frac{E_C}{c_0^2} 
\label{Def-delta-m}\\
\Delta a_C &=& \frac{\delta m_P}{m} g_0 
\label{Def-Delta-a}
\end{eqnarray}

Equations (\ref{Var-N-1}) and (\ref{Var-N-Def}) to (\ref{Def-Delta-a})
are the main result of this letter. They show that a breakdown
of the weak equivalence principle by electromagnetic interactions and
conservation of charge are not independent.  Furthermore, the
conservation of $Q^*$ implies that there is a current of neutral
particles, carrying out particle number, from the decay of the charged
ones. 

{\bf 2:} {\it Extension to superstring theories:} The main result
obtained in the previous section can be extended to some cases of
superstring theories, namely those with a massless dilaton. Let us
concentrate in the matter action of the model proposed in
ref.\cite{DP}
\begin{eqnarray}
S_m &=& - \int d^4 x \sqrt{g}\left[-\psi \gamma^{\mu}\left(\partial_{\mu} 
- i A_{\mu}\right)\psi\right] \nonumber \\
& & - \int d^4 x \sqrt{g} \frac{1}{4} k B_f \left(\phi\right) F^{\mu
\nu} 
F_{\mu \nu}
\label{DP}
\end{eqnarray}

The third term of last equation accounts for the electromagnetic field
contribution with $\epsilon = \mu^{-1}$. Furthermore, we can identify
the second term with the coupling between matter and electromagnetism,
and the first one with the kinetic contribution with $H = T = 1$.

Thus, eq.\ref{Var-N-1} holds and since 
the charge
measured in the free falling system is $e^*$, we can 
write the
following equation:
\begin{eqnarray}
\frac{\dot \alpha}{\alpha} = 2 \frac{\dot e^*}{e*} = -2 
\frac{\dot N}{N}
\label{rel-alpha-N}
\end{eqnarray}

	It can be shown \cite{DP} that in this model the following
relation exists between the the anomalous acceleration $\Delta a_C$
and the cosmological variation of $\alpha$:
 \begin{equation}
\frac{\Delta a_C}{a} \sim 10^{-2} \frac{\dot \alpha}{\alpha H_0} =
\frac{2 \times 10^{-2}}{H_0} \frac{\dot N}{N} 
\label{damourpol}
\end{equation}
which is peculiar to this model.

3: \emph{Bekenstein-like theories}

In order to study the fine structure constant variability,
Bekenstein \cite{Bekenstein82} proposed a theoretical framework based
on very general assumptions. In this context, every particle charge
can be expressed in the form $e=e_0 \varphi \left(\vec x,t)\right)$
where $\varphi \left(\vec x,t)\right)$ is a scalar field and the
matter action of a system of particles can be written as follows:

\begin{eqnarray}
S_{NG}&=&- \frac{1}{16 \pi} \int \varphi^{-2} F^{\mu \nu} F_{\mu \nu}
d^4x \\ 
& & + \sum_i \int \left[m c^2 +\frac{e_0}{c} u^{\mu} A_{\mu}\right]
\gamma^{-1} \delta^3\left(x^i-x^i(\tau)\right) d^4x  \nonumber 
\end{eqnarray} 
where $A_{\mu}$ is $\varphi$ times the gauge field as defined by
Bekenstein.

Thus, we can identify the first term with the electromagnetic
contribution in the $TH\epsilon \mu$ formalism with $\epsilon =
\mu^{-1}= \varphi^{-2}$ and the second term with the matter and
coupling between matter and electromagnetism with $T=H=1$.  For $N$
identical particles we have:
\begin{equation}
\frac{\dot N}{N}=-\frac{\dot \varphi}{\varphi}
= -\frac{1}{2}\frac{\dot \alpha}{\alpha}
\end{equation}
where $\alpha$ is the fine structure constant. It is easy to show that
a relation similar to eq.\ref{damourpol} holds in this model. Using
the results in ref.\cite{Bekenstein82} we obtain:
\begin{equation}
\frac{\Delta a_C}{a} \sim 2 \times 10^{-3} \frac{\dot \alpha}{\alpha H_0} =
\frac{4 \times 10^{-3}}{H_0} \frac{\dot N}{N} 
\label{Beck-eta-alpha}
\end{equation}

4: \emph{Comparison with experiments}

There have been many experiments to put bounds on processes that
change charge discontinuously, such as the dissapearance of electrons
\cite{Okun92}. Thus, we can use these results to put bounds on
$\alpha$ variation and can use the relation between $\alpha$
variation and $\Delta a/a$ of eq.\ref{damourpol}  to put bounds on
the breakdown of the equivalence principle.  Similar results, somewhat
stronger, hold for Beckenstein-like models, improving the limits
established in ref.\cite{Bekenstein82} on violation of
EEP. Results are shown in Table \ref{ncqtab}.

When we use our relations (\ref{Var-N-Def}) to (\ref{Def-Delta-a}) we
have
\begin{equation} 
\frac{\Delta a_C}{a} = 8 \frac{\dot N}{N} \frac{E_c}{m c_0^2}
\frac{c_0^2}{g_0 \cdot v}.
\end{equation}
For the fall towards the Virgo Cluster we estimate $v/c_
0\simeq 10^{-3}$, 
$g_0/c_0^2\simeq 10^{-16} m^{-1}$ and typically $E_c/m c_
0^2\simeq 10^{-3}$
and we obtain
\begin{equation}
\frac{\Delta a_C}{a}= 3 \times 10^{14} y \frac{\dot N}{N} 
\leq 10^{-12}
\end{equation}
which is a much weaker bound that the one using Damour and Polyakov
model relationship and even more weak than the limit obtained from
the Bekenstein model relationship.

This can be understood as follows: the bound from expression
(\ref{Var-N-Def}) comes from the anomalous coupling of the
electromagnetic energy with gravity, while the bound from expression
(\ref{damourpol}) comes from the dilaton exchange mechanism as used in
\cite{DP}, which is a much more strong effect than the electromagnetic
one. On the other hand, expression (\ref{Beck-eta-alpha}) comes from
the close link between the gradient of the gravitational potential and
the gradient of $\alpha$ in Bekenstein theory, which we do not
consider in our adiabatic $TH\epsilon\mu$ treatment.


We see then that there is a deep connection between charge
non-con\-ser\-vat\-ion and violation of university of free fall for a
wide class of theories, namely those that can be written in the
$TH\epsilon\mu$ form. The connection, as expressed in
(\ref{Var-N-Def}), is considerably general, and provides a link
between any electromagnetic violation of EEP and non conservation of
charge. The corresponding bounds on WEP are comparable to present day
values. The connection (\ref{damourpol}) is more specific from
dilaton-type theories, a special case of $TH\epsilon\mu$ theories,
which are those that provide the mechanism considered in \cite{DP}.
In this case the bounds obtained are even lower than proposed future
direct tests of WEP \cite{Will98}. Consequently these future tests
still deserve much attention, though they may add not too much new
information as regards to dilaton-type gravitational theories, as long
as the local coupling of the dilaton field can be neglected.

\begin{table}
\begin{tabular}{lllllll}
\hline
Process& Ref. &  $\tau$ (y)&  $\mid\frac{\dot\alpha}{\alpha}\mid (y^{-1})$&$\mid\frac{\Delta a}{a}\mid_{DP}$& $\mid\frac{\Delta a}{a}\mid_{Beck}$ &$\mid\frac{\Delta a}{a}\mid_{TH\epsilon\mu}$ \\
\hline
$^{71}{\rm Ga} \rightarrow ^{71}{\rm Ge}$ & \cite{NBG96} 
& $ 4\times 10^{26}$ & $ 6 \times 10^{-27}$ & $ 
10^{-18}$ &$2 \times 10^{-19}$& $ 2 \times 10^{-13}$ \\ 
$ e \rightarrow \nu_e \gamma $ & \cite{Avign86} & $ 
2   \times 10^{25}$
&   $  8 \times 10^{-26}$ & $ 10^{-17}$ &$2.5 \times 10^{-18}$&  $  2 \times 10^{-11}$\\ 
$ e \rightarrow {\rm any}$& \cite{Reusser91} & $ 3 \times10^{23}$  & $ 8 \times 10^{-24}$ &  $ 10^{-15}$  & $4 \times 10^{-16}$&
$ 2 \times 10^{-9}$ \\ 

$ e \rightarrow \nu_e \gamma $ & \cite{Belli00} & $ 
2   \times 10^{26}$
&   $  5 \times 10^{-27}$ & $ 2 \times 10^{-18}$ &$3 \times 10^{-19}$&  $  2 \times 10^{-12}$\\ 
$ e \rightarrow {\rm any}$& \cite{Belli99} & $ 2 \times10^{24}$  & $ 10^{-24}$ &  $ 5 \times 10^{-16}$  & $3 \times 10^{-17}$&
$ 2 \times 10^{-10}$ \\ 
\hline 
\end{tabular}
\caption{Results. The 
columns show the process considered, the corresponding 
references, the 
observational data, the limits on the time-variation of the fine structure
constant and the 
bounds for the breakdown of the equivalence principle obtained from Damour and Polyakov-like theories, Bekenstein-like models and the general relationship derived in this paper.}
\label{ncqtab}
\end{table}

\end{document}